\documentclass[a4paper,fleqn,usenatbib]{mnras}

\usepackage[T1]{fontenc}
\usepackage{ae,aecompl}

\usepackage{graphicx}	
\usepackage{amsmath}	
\usepackage{amssymb}	
\usepackage{wasysym}

\title[Chemical Tracers of Pre-Brown Dwarf Cores]{Chemical Tracers of Pre-Brown Dwarf Cores Formed Through Turbulent Fragmentation}

\author[J. Holdship and S. Viti]{Jonathan Holdship$^{1}$\thanks{E-mail:jrh@star.ucl.ac.uk} and Serena Viti$^{1}$
\\
$^{1}$Department of Physics and Astronomy, University College London, Gower Street, London, WC1E 6BT
}

\date{Accepted XXX. Received YYY; in original form ZZZ}

\pubyear{2015}

\begin{document}
\label{firstpage}
\pagerange{\pageref{firstpage}--\pageref{lastpage}}
\maketitle

\begin{abstract}
A gas-grain time dependent chemical code, UCL\_CHEM, has been used to investigate the possibility of using chemical tracers to differentiate between the possible formation mechanisms of brown dwarfs. In this work, we model the formation of a pre-brown dwarf core through turbulent fragmentation by following the depth-dependent chemistry in a molecular cloud through the step change in density associated with an isothermal shock and the subsequent freefall collapse once a bound core is produced. Trends in the fractional abundance of molecules commonly observed in star forming cores are then explored to find a diagnostic for identifying brown dwarf mass cores formed through turbulence. We find that the cores produced by our models would be bright in CO and NH$_3$ but not in HCO$^+$. This differentiates them from models using purely freefall collapse as such models produce cores that would have detectable transitions from all three molecules.
\end{abstract}




\section{Introduction}
A brown dwarf is a substellar object with a mass less than 0.075 M$_{\astrosun}$; the H-burning limit. Since their discovery in 1995 \citep{rebolo1995}, the origin of brown dwarfs has challenged simple theories of star formation. Various theories have been put forward to explain how objects much smaller than the average Jeans mass in a molecular cloud can become gravitationally bound and unstable to collapse. There is expected to be some overlap between the lowest mass brown dwarfs and highest mass planets. Precisely how brown dwarfs form may therefore be the main criterion for categorizing them.\par
For example, if brown dwarfs form by condensing out of the outer regions of a massive prestellar disk \citep{rice2003}, then they would be hard to separate from giant planets. Other formation theories include ejection from a common envelope \citep{reipurth2001} in which several protostars fragmenting from a core accrete competitively leaving one at much lower mass than the others. In simulations of three or more of these mutually bound objects, the lowest mass companion is regularly ejected. If this happens whilst the object is substellar, it will remain so as it will no longer have an envelope to accrete from.\par
Another way brown dwarfs could form is through the photo-erosion of a prestellar core \citep{hester1996}. If a star-forming core is overrun by a HII region, its outer layers will be ionized potentially liberating enough mass to leave an object with a total mass below the H-burning limit.\par
Finally, the theory considered in this work is that of turbulent fragmentation \citep{padoan2002,hennebelle2008}. In these theories, bound cores are formed by the collision of supersonic flows within molecular clouds. The resulting isothermal shocks produce dense cores of gas which may collapse to form stars. The range of core masses produced depend on the cloud's turbulent power spectrum and spans masses from the smallest brown dwarfs to the most massive stars. Both of the referenced works produce a core mass function consistent with theory and observation, implying that a universal star formation process could be responsible for all masses of stellar objects.\par
Differentiating between these theories requires observations of brown dwarfs in the earliest stages of their evolution and few of these have been found. In principle, the physical characteristics of more evolved brown dwarfs could be used to identify the formation process. However, this has its difficulties. For example, a brown dwarf having a small disc could imply formation through ejection as the disc is expected to be truncated by the gravitational interaction. Unfortunately, simulations by \citet{bate2012} have shown that whilst most discs are truncated to \textless40 AU, discs can extend as far as 120 AU. This means that despite recent ALMA observations \citep{ricci2014} of three brown dwarfs which found all had disks \textgreater70 AU, ejection cannot be ruled out though it does seem disfavoured.\par	
Several candidate proto-brown dwarfs have been found in the follow up of the Very Low Luminosity Objects (VeLLOs) first discovered by Spitzer \citep{kauffmann2005}. A recent candidate from SMA observations by \citet{palau2014}, is a good example of a brown dwarf mass object in the earliest stages of its evolution. Its mass is very likely in the brown dwarf range and it has many features that are found in low mass stars at the same stage. The similarity of these `proto-brown dwarfs' to low mass protostars seems to imply that both sets of objects form in the same way. This view favours more universal theories of star formation such as the turbulent fragmentation scenario as they would produce both low mass stars and brown dwarfs through the same mechanism.\par
Whilst using physical clues to find the main route for brown dwarf formation has proven difficult, it may be possible to differentiate the models chemically. The molecular abundances in a proto-brown dwarf formed from material in the disk of a massive star is likely to be different to that of a proto-brown dwarf formed in a shocked region of a molecular cloud.\par
In this paper, we present an adaptation of the chemical model UCL\_CHEM \citep{viti2004} that simulates the formation of a brown dwarf through turbulent fragmentation. The aim is to present molecular tracers that could be used to identify a pre-brown dwarf core that has formed through shock compression. Chemical abundances are highly sensitive to the cloud density so the fast increase that accompanies a shock in a molecular cloud should give rise to certain molecules having abundances incompatible with a less violent density increase such as a freefall collapse. In Section~\ref{sec:model}, the chemical code and physical model are described. In Section~\ref{sec:results}, the effect of changing model parameters on the fractional abundance of molecules is explored along with the variation of species abundance with radial position. In Section~\ref{observer}, observational tests of the models are suggested and the models are compared to observations. Finally, a summary is given in Section~\ref{sec:summary}.

\section{The Model}
\label{sec:model}

The chemical code UCL\_CHEM \citep{viti2004} was adapted to model the turbulent fragmentation and collapse of a low mass core. In this work, the chemical model follows the abundances of 160 gas phase species and 54 grain surface species through a network of 2413 reactions for a box of gas in a collapsing molecular cloud. By running several of these boxes at points along a line of sight in a molecular cloud, a one dimensional model can be built up. The physical model comprises three stages, in which the density of the gas changes in different ways. Density is given here as the total number of hydrogen nuclei per cubic centimetre.\par
In stage one, the gas density increases in freefall from n$_0$=10$^2$cm$^{-3}$ up to a final density typical of a molecular cloud, usually 10$^4$cm$^{-3}$. Whilst the density is then held at this value, the model continues to run for a further  1 or 10 Myr. This allows the age of the cloud to be varied. The temperature is set to 10 K or 30 K in this stage and held constant throughout. The purpose of stage one is to produce a self consistent molecular cloud for the next two stages rather than assuming the conditions and abundances of a molecular cloud.\par
In stage two, the formation of cores by turbulent fragmentation is simulated. We consider a bound core that is formed as supersonic flows within a molecular cloud collide. In this process, the centre of the core forms first and grows outwards as gas is deposited onto it. The radius of the core would grow at a rate given by the velocity of the flow. To model this, the density and chemistry of 10 points are followed, spaced evenly between the centre and edge of the core. The innermost point (i$_{point}$=0) is immediately increased to the post-shock density and each point afterwards is increased to the same density at a time given by its distance from the centre and the shock speed, as in Equation~\ref{eq:tshock}.

\begin{equation}
\label{eq:tshock}
t_{shock} = \frac{i_{point}}{N_{points}-1}\left(\frac{r_{shock}}{M_{s}C_{s}}\right)
\end{equation}

Where $r_{shock}$ is the radius of the final, post-shock core and $i_{point}$ is a counter ranging from 0 to ($N_{points}-1$). The post-shock density, $n_{shock}$, is taken to be the critical density of a Bonnor-Ebert sphere of the mass chosen for the model and is given by Equation~\ref{eq:dens} taken from \citet*{padoan2004}. The final density is proportional to $M^2$ in an isothermal shock, thus the Mach speed of the shock is set by the required post shock density in this model. A sound speed of $C_s$=0.2 kms$^{-1}$ is assumed.\par

\begin{equation}
\label{eq:dens}
n_{shock} \sim 10^3cm^{-3}\left(\frac{M_{BD}}{3.3M_{\astrosun}}\right)^{-2}\left(\frac{T}{10 K}\right)^3
\end{equation}
 In the third stage of density increase, each point begins to collapse according to the same freefall equation as stage one immediately after reaching $n_{shock}$. This means that the innermost points in the core are already collapsing by the time the shock reaches the outer points. This collapse is to a maximum density of 10$^8$cm$^{-3}$, beyond which three body reactions become non-negligible. The chemical network used does not include these reactions.\par

These shocks were considered to be isothermal, the temperature set in stage one (typically 10 K) is kept constant. A short period of increased temperature immediately following the shock was included to test the effect this would have. A typical post-shock temperature for a strong shock at the shock speeds used in these models is 125 K \citep{draine2011}. Including a temperature increase up to 500 K did not significantly alter the results and so the period of temperature increase was subsequently left out of the model.\par
The fractional abundances of molecular species and the density of each point are averaged for each time step giving the average density and species' fractional abundances for a line extending from the centre to the edge of a core. Whilst UCL\_CHEM keeps track of a total of 214 species, we focus on five molecules in particular: CO, HCO$^+$, CS, NH$_3$ and N$_2$H$^+$. These molecules are chosen as they are amongst the most commonly observed molecules in molecular clouds (see Section~\ref{sec:comparison}). \par

To explore the effect of varying initial conditions, eight sets of parameters were used in stage one to produce different starting points for stages two and three. Table~\ref{tab:models} lists the sets of conditions reached in stage one. Throughout the paper, the models and figures are labelled according to the following nomenclature: N-M$_{BD}$. N refers to the model number in Table~\ref{tab:models}, giving the parameters used in stage one for that model. M$_{BD}$ is the mass of the core chosen to be modelled. For example, 1-0.07 refers to a stage one run at 10 K to produce a cloud of age 1 Myr limited to a density of 10$^4$cm$^{-3}$ which is subsequently used for the initial conditions to simulate the creation of a 0.07 M$_{\astrosun}$ core.

\begin{table}
\centering
\caption{List of final conditions reached by stage one models. Each stage two model uses the final output of a stage one model as its initial conditions.}
\begin{tabular}{cccc}
\hline
\textbf{Model} & \textbf{Temp/ K} & \textbf{Final Density/ cm$^{-3}$}  & \textbf{Time /Myr}\\
\hline
1&10&10$^4$&1\\
2&10&10$^5$&1\\
3&30&10$^4$&1\\
4&30&10$^5$&1\\
5&10&10$^4$&10\\
6&10&10$^5$&10\\
7&30&10$^4$&10\\
8&30&10$^5$&10\\
\hline
\end{tabular}
\label{tab:models}
\end{table}

\section{Results}
\label{sec:results}
\subsection{Sensitivity to Initial Conditions}
It is important to know how sensitive the model is to the initial conditions of the cloud. To this end, eight variations of stage one were explored each with varying cloud age, density and temperature. \par
The age of the cloud, taken as the amount of time elapsed since the cloud reached its final stage one density, was tested at 1 Myr and 10 Myr. Changing the age of the cloud leads to larger freeze out for older clouds and thus lower initial gas phase molecular abundances at stage two. However, these changes were less than an order of magnitude on average and are almost completely lost in the post-shock gas. \par
The density of the cloud has a larger effect. Not only are the initial abundances for stage two different, but the shock itself is also less violent. The density increase required to create a bound core of the chosen mass is smaller for higher density clouds and thus the shock is slower. However, the differences in molecular abundances diminish throughout stage two and are lost in the freefall collapse after the shock. \par
Of the three cloud properties varied, temperature has the largest effect. It affects chemistry by changing the rates of reactions and also changes the shock speed. Higher temperature gas must be more dense to be bound and so higher temperature models must have faster shocks to produce higher densities. During the shock, molecular abundances vary by more than two orders of magnitude on average. However, before and after the shock abundances differ by less than an order of magnitude and so many of the results presented in this paper are unaffected. This is discussed further in Section~\ref{sec:predict}.

\subsection{Stratification}
Comparisons to the core as a whole could well be enough to match observations to the models and determine whether they could be formed through turbulent fragmentation. However, modern instruments could resolve a core, allowing for characterization of its density and temperature structure. For example, the full capability of ALMA will allow angular resolutions of 0.015", allowing scales as small as 2 AU to be resolved at distances of 140pc; the distance to the Taurus molecular cloud. It is therefore possible to ask whether the abundance of certain molecules would be expected to be higher in the centre of the core or at the edge.\par
\begin{figure*}
\centering
\includegraphics[width=0.8\textwidth]{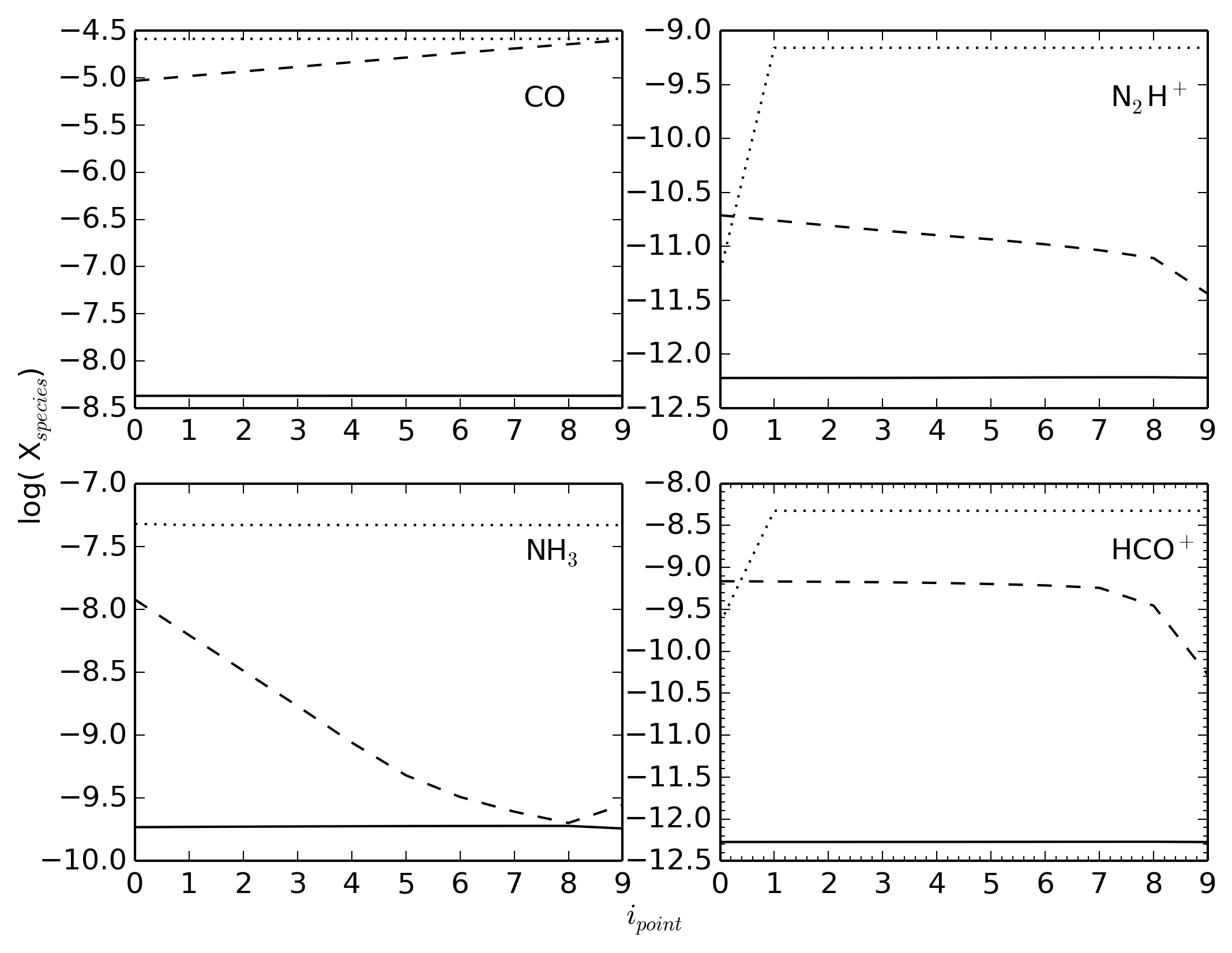}
\caption{Fractional abundances with distance from core centre for four molecules, the model shown is a 0.07 M${\astrosun}$ core formed from the first set of initial conditions. The x-axis shows the point number, with the centre of the core being point 0 and the edge being 9 for a total of 10 points distributed along the core radius. The dotted lines show the start of stage two where only the first point has had a density increase, the dashed line shows the end of the shock passage when the final point has been reached and the solid line shows 0.1 Myr after the shock.}
\label{fig:strat}
\end{figure*}

The models are run as a series of single point models along a radial line from the centre to the edge of a core, allowing stratification to be seen by plotting fractional abundance against radial point position for given times. Figure~\ref{fig:strat} shows the abundances of four molecules at varying times for the 1-0.07 model. Of the five molecules examined in this paper, four have been plotted as CS follows the behaviour of CO. \par
The  dashed lines in Figure~\ref{fig:strat} show the fractional abundance of molecules just after the shock reaches the final point; when the core is considered to be just formed. At this time, there can be large differences between the core's edge and centre. NH$_3$ in particular is over an order of magnitude more abundant in the centre than at the edge. However, the solid lines show 0.1 Myr later and there are no differences in fractional abundances between points at this time. Given that the shock period is relatively short compared to the lifetime of the core, it is unlikely that a core will be observed mid formation and so no stratification will be observed.

\section{Observational Tests}
\label{observer}
\subsection{Observational Predictions}
\label{sec:predict}

The models produced here may be best used by comparing them in detail to individual pre-brown dwarf cores. However, a number of general conclusions can be drawn from the models. If the majority of pre-brown dwarf cores are formed through turbulent fragmentation then it is expected that they will be consistent with our predictions.\par
In order to qualify whether our predictions can lead to observables, the radiative transfer code RADEX \citep{vandertak2007} was used to give an indication of what might be observable in pre-brown dwarf cores considering the fractional abundances given by the models. RADEX was run repeatedly using the results of the models at different time steps to produce the line intensities for several molecular transitions as a function of time. For each time step, the density was halved to give the number density of H$_2$ molecules, as the vast majority of H nuclei are in H$_2$ for all models. A kinetic temperature of 10 K is used, in line with the model temperature. The line width is assumed to be turbulence dominated and taken to be 1 kms$^{-1}$. RADEX also requires a column density which was calculated by multiplying the density of each species by the radius of the core.\par
By considering the brightest line emitted by each molecule that would be visible using a ground based telescope and applying a lower limit for visibility, several key diagnostics of a core formed through turbulent fragmentation were found. The capabilities of ALMA and the VLA were used as a guide. Of course, the results given assume most of the flux is recovered which may not be the case.. No adjustment for beamsize has been made as both arrays have synthesized beamwidths smaller than the typical angular size of these cores; $\sim$10"  at 140 pc. T$_{MB}$ = 100 mK has been used as the lower limit for a detectable line (see Section~\ref{sec:comparison}). The results discussed in this section are based on models using the first set of initial conditions: n$_H$=10$^{4}$cm$^{-3}$, T=10K and an age of 6 Myr. \par
\begin{figure*}
\centering
\includegraphics[width=0.9\textwidth]{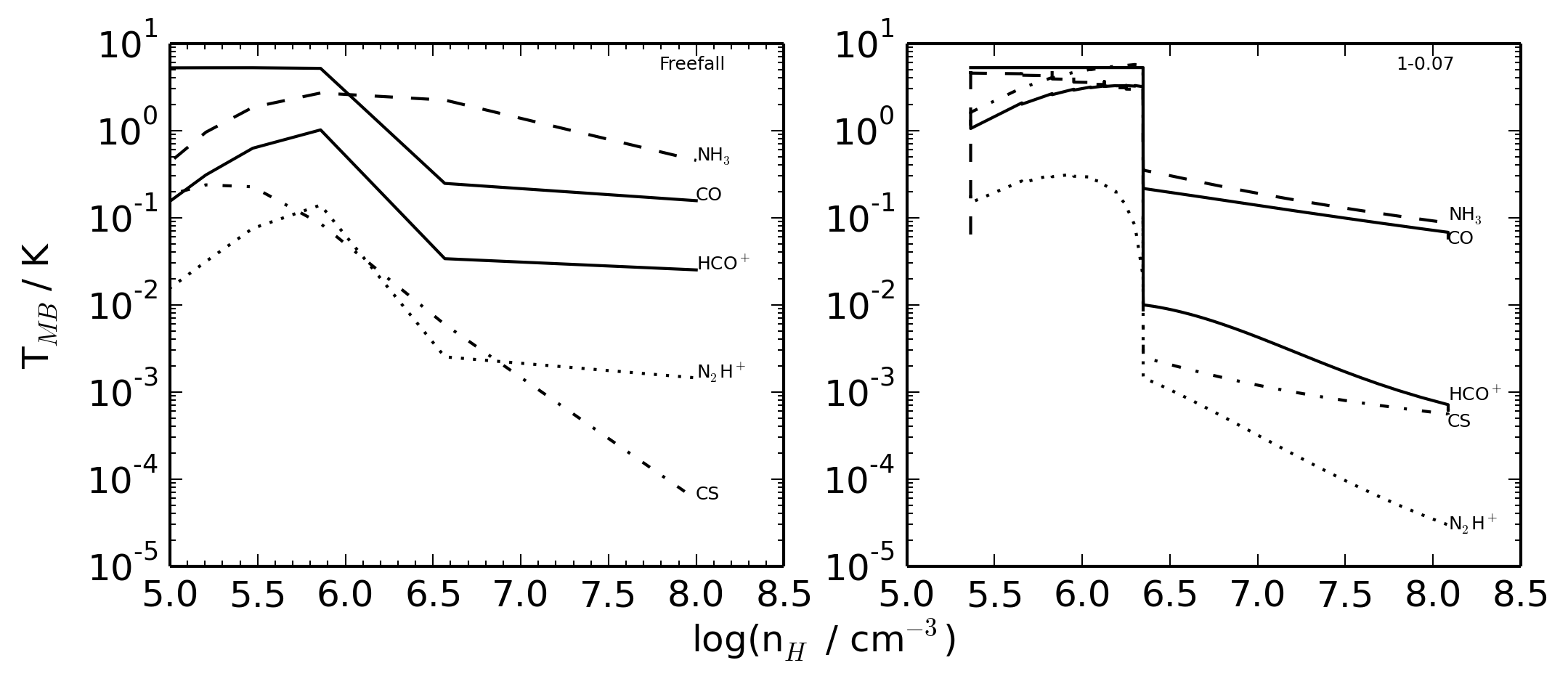}
\caption{Brightness temperature expected for the five transitions discussed in Section~\ref{sec:predict}. On the left, a freefall model is shown and on the right a turbulent fragmentation model in which a 0.07 M${\astrosun}$ core is formed from a cloud using the first set of initial conditions. A clear gap is apparent in the turbulent fragmentation model with CO and NH$_3$ being much brighter than other molecules.}
\label{fig:flux}
\end{figure*}

For bound cores above 0.04 M${\astrosun}$, detection of the CO 2-1 and NH$_3$ 1$_{(1,0)}$-1$_{(1,0)}$ transitions are expected. They are difficult to differentiate from a core formed through freefall using the 100 mK cut off. However, the behaviour of the HCO$^+$ 3-2 transition is very different between turbulent fragmentation model and freefall. HCO$^+$ is an order of magnitude brighter in freefall cores than in the turbulent fragmentation models. A comparison between a  0.07 M${\astrosun}$ and equivalent freefall model is shown in Figure~\ref{fig:flux}.\par
This is not the case for an extremely low mass core (\textless 0.04 M${\astrosun}$). Such cores are extremely dense and are formed very quickly, giving bright lines in CS 3-2 and N$_2$H$^+$ 1$_{(2,3)}$ - 0$_{(1,2)}$ which are not apparent in higher mass cores or cores formed through freefall. Cores are unlikely to be detected during the shock as this is a very short lived stage. However, an unusually bright CS 3-2 line would be an indicator of a shock in progress.\par

Table~\ref{table:observe} shows which of the molecules considered in this paper are observable in cores at densities of n$_H>$10$^7$cm$^{-3}$. The models shown use Model 1 conditions. 30 K models have much higher $n_{shock}$ values and so molecules remain detectable at higher densities. All five molecules are detectable up to at least n$_H=$10$^8$cm$^{-3}$ for the 0.04 and 0.01 M$_{\astrosun}$ models at 30 K.\par

\begin{table}
\centering
\caption{Observable molecules for different models at n$_H>$10$^7$cm$^{-3}$. Where an observable molecule is considered to be one that has a transition which would emit with a brightness temperature greater than 100 mK.  All models use the first set of initial conditions. A + indicates an observable molecule.}
\begin{tabular}{cccccc}
\hline
Molecule&0.01 M$_{\astrosun}$&0.04 M$_{\astrosun}$ &0.07 M$_{\astrosun}$&0.10 M$_{\astrosun}$&Freefall\\
\hline
CO&+&+&+&+&+\\
HCO$^+$&+&&&&\\
CS&+&&&&\\
NH$_3$&+&+&+&+&+\\
N$_2$H$^+$&+&&&&\\
\hline
\end{tabular}
\label{table:observe}
\end{table}

\subsection{Comparison to Observations}
\label{sec:comparison}
\begin{figure}
\centering
\includegraphics[width=0.5\textwidth]{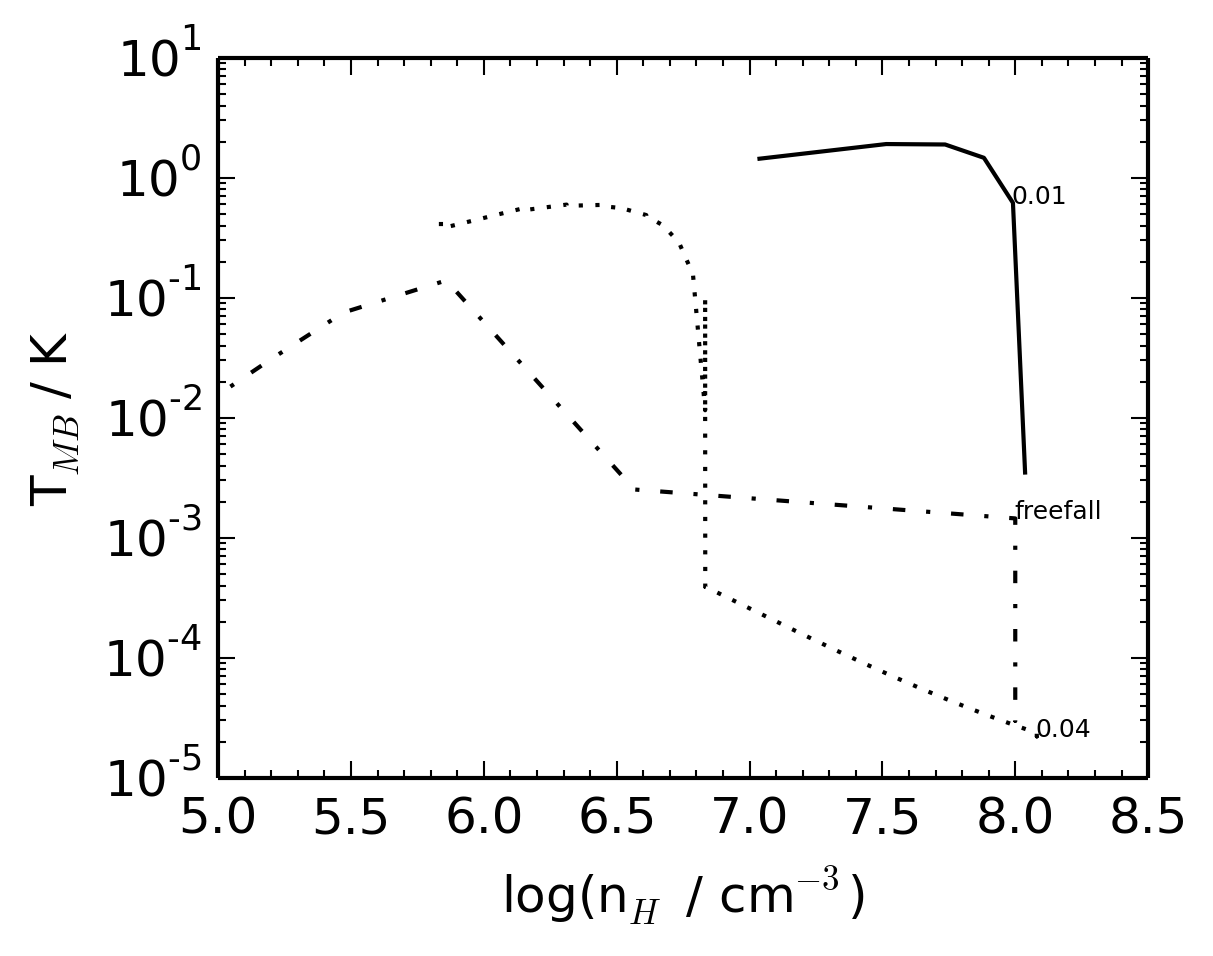}
\caption{Brightness temperature of the N$_2$H$^+$ 1$_{0,1}$-0$_{1,2}$ transition for three models; 0.01 M$_{\astrosun}$,0.04 M$_{\astrosun}$ and freefall. Only the lowest mass turbulent fragmentation model can produce a bright line for this transition at n$_H$=7.5x10$^7$cm$^{-3}$; the density calculated by \citet{andre2012}.}
\label{fig:andre}
\end{figure}

Only one pre-brown dwarf core has been detected so far. \citet{andre2012} presented observations of a core of mass less than 0.03 M$_{\astrosun}$ with a good detection in the 3.2 mm continuum and the N$_2$H$^+$ 1$_{0,1}$-0$_{1,2}$ line. The core is estimated to be at n$_H$\textgreater7.5$\times$10$^7$cm$^{-3}$. Figure~\ref{fig:andre} shows the RADEX output for this transition for the models as a function of density. Whilst a N$_2$H$^+$ detection at this density would not fit with freefall or higher mass models it would fit with low mass models where the density has been greatly increased by the shock. This is by no means conclusive but it does show that there are measured abundances in very low mass cores that do not fit with simple freefall models and can be fitted by assuming a fast density increase due to a shock.\par

The turbulent fragmentation process is not brown dwarf specific, therefore prestellar cores in the low mass star range could be used to test the model until more pre-brown dwarf cores are discovered. \citet{seo2015} used a source finding algorithm with NH$_3$ observations of the L1495-B218 filaments in the Taurus molecular cloud and found many low mass cores which are potentially prestellar cores. In their work Seo et al. present core masses estimated from dust emission and from NH$_3$ column densities using an assumed fractional abundance of NH$_3$. Using these to calculate their measured mass of NH$_3$ and assuming the mass from dust is the total core mass, we have arrived at NH$_3$ fractional abundances for their cores. Table~\ref{table:cores} shows a comparison between those NH$_3$ abundance estimates and the models. The models show poor fits for the brown dwarf mass cores. However, it is worth noting that none of these cores are gravitationally bound according the Virial masses calculated by \citet{seo2015}. The models in this paper use the mass to set $n_{shock}$ assuming that a bound core will be produced, therefore an unbound core is not expected to be a good match to the models. For  this reason models were run with higher mass cores to compare to bound objects in the paper. For the higher mass cores in Table~\ref{table:cores}, the fractional abundance of NH$_3$ is much closer and is a good match considering that without density measurements, the abundances can change by a factor of a few depending on which density value is chosen. The model abundances chosen are during the post-shock period where many molecules maintain constant fractional abundances for close to a million years. This is the most likely stage in which to view the cores as it is by far the longest.\par
\begin{table}
\centering
\caption{List of cores from \citet{seo2015} with model equivalents. Fractional abundances given in units of 10$^{-9}$, those from the model are taken to be the average abundance for the post shock core.}
\begin{tabular}{cccc}
\hline
Mass/ M$_{\astrosun}$ & Virial Mass/ M$_{\astrosun}$ & X(NH$_3$)$_{observed}$ & X(NH$_3$)$_{model}$\\
\hline
0.04&0.31&1.13&0.07\\
0.07&0.22&1.07&0.32\\
0.10&0.58&1.8&0.5\\
0.88&0.77&3.65&1.5\\
0.95&0.98&1.35&2.5\\
0.99&1.09&1.00&1.0\\
\hline
\end{tabular}
\label{table:cores}
\end{table}
For further comparison, the models were fitted to data on dense cores in the Lupus molecular cloud, presented by \citet{benedettini2012}, which could be prestellar cores. Whilst the density and mass of these cores is unknown, they do calculate the column densities of up to 5 species for each core. The three cores (Lupus1 C4,C6 and C7) with the most column density data were chosen and the column densities were used to compare the ratios of CH$_3$OH,HC$_3$N, N$_2$H$^+$ and NH$_3$ with CS to the same ratios from our models. For each model, the best fit density was found through a $\chi^2$ test and used to evaluate which model and density would best fit the cores. This test was done for all the initial conditions and brown dwarf masses; none were considered to be a good fit. This seems to imply that, whether they are gravitationally bound or not, the objects presented by \citet{benedettini2012} are not high density cores formed through turbulent fragmentation.

\section{Summary}
\label{sec:summary}
A gas-grain time dependent chemical code, UCL\_CHEM, has been adapted to investigate the formation of brown dwarfs through turbulent fragmentation. To do this a multi-point model has been used to emulate a 1D model of a shock travelling through a cloud. The resulting chemistry is expected to be representative of a prestellar core formed by the collision of turbulent flows. Without more observations to compare to, it is difficult to say how reliable the models are. However, certain aspects are promising as they essentially reproduce the results of freefall models for low mass stars and show behaviour that is incompatible with freefall in brown dwarf cores where the effects of turbulent compression would be strongest. For example, in an extremely low mass brown dwarf, the N$_2$H$^+$ abundance is three orders of magnitude higher than in an equivalent freefall model.\par
The model is not strongly sensitive to the initial conditions investigated, though changes in the gas temperature do affect the results. In principle, temperature measurements may included when fitting with the models. However, despite this sensitivity, the models produce molecular abundances which are incompatible with freefall models for a wide range of initial conditions. This has allowed us to provide general observational tests that could mark a core as worth following up with more observations. These include the detection of CO and NH$_3$ in a low mass core without the detection of HCO$^+$  which would be expected from freefall. Whilst the existence of pre brown dwarf cores does in itself reduce the possible formation routes, the possibility of matching the chemistry to a specific formation model would add to the body of evidence available for determining how very low mass objects form.\par
Modern telescopes offer the opportunity to look for these small, dark objects and observe them with enough resolution to obtain data on their molecular abundances. One such object has already been observed, implying that more could be found in the near future. This model could then be used to add to the evidence available when considering formation processes.\par
Introducing the shock to the model produces many changes to the chemistry when compared to freefall models. The largest effects of the shock are in maintaining high abundances of molecules at high core densities where they would normally have frozen out during a freefall collapse. An example is CS which is undetectable at X$_{CS}$ \textless 10$^{-12}$ in a core of density n$_H$=10$^7$cm$^{-3}$ for freefall models but has a fractional abundance of X$_{CS}$=10$^{-8}$ for the lowest mass brown dwarf model. Similarly, CS freezes out more quickly than CO during a freefall, leading to a larger ratio at higher densities during a freefall collapse. However, this ratio does not increase during the shock as both species remain at roughly constant fractional abundance. This flattening in the CS trends also means that it is never less abundant than N$_2$H$^+$ in the shock models whilst it freezes out more quickly than N$_2$H$^+$ in freefall and is generally less abundant.\par
Due to the chemical differences between the freefall and turbulent fragmentation models, it is reasonable to expect that the most likely formation process responsible for the formation of a particular brown dwarf could be found by an extensive spectral survey of pre-brown dwarf cores. This could include measurements of fractional abundances to compare to the models presented here as well as investigations into line profiles to reveal the importance of turbulence in regions forming brown dwarfs.
\section*{Acknowledgment}
We are grateful to the referees for their comments on a previous version of this paper which have led to a much stronger manuscript.




\bibliographystyle{mnras}
\bibliography{turbfrag} 

\bsp	
\label{lastpage}
\end{document}